\documentclass[useAMS,usenatbib]{mn2e}
\usepackage{amsmath}
\usepackage{amsfonts}
\usepackage{amssymb}
\usepackage{graphicx}
\usepackage{aas_macros}
\usepackage{ifthen}
\usepackage{color}

\def\blackandwhite{false}

\ifthenelse{\equal{\blackandwhite}{true}}
{
  \newcommand{\fig}[1]{f#1_bw}
  \newcommand{\darkred}[1]{\textcolor[gray]{0.6}{#1}}
}
{
  \newcommand{\fig}[1]{f#1}
  \newcommand{\darkred}[1]{\textcolor[rgb]{0.75,0,0}{#1}}
}

\newcommand\plotone[2]{%
 \centering
 \leavevmode
 \columnwidth=#2\columnwidth
 \includegraphics[width={\columnwidth}]{#1}%
}%

\newcommand{\figref}[1]{Figure~\ref{#1}}
\newcommand{\eqnref}[1]{Eq.~\eqref{#1}}
\newcommand{\secref}[1]{Section~\ref{#1}}

\newcommand{\tabref}[1]{Table~\ref{#1}}
\newcommand{\mean}[1]{\ensuremath{\langle #1 \rangle}}
\newcommand{\e}[1]{\ensuremath{\times 10^{#1}}}

\renewcommand{\vec}[1]{\ensuremath{\boldsymbol{#1}}}
\newcommand{\mat}{\mathbf}

\newcommand{\Scovar}{\ensuremath{\widehat{\mat\Sigma}}}
\newcommand{\Scovarmod}{\ensuremath{\widetilde{\mat\Sigma}}}

\title[A Sampling Strategy for High-Dimensional Spaces]{A Sampling Strategy for High-Dimensional Spaces Applied to Free-Form Gravitational Lensing}

\author[Mario Lubini and Jonathan Coles]%
{Mario Lubini\thanks{E-mail: lubini@physik.uzh.ch} and Jonathan Coles\thanks{E-mail: jonathan@physik.uzh.ch}\\%
Institut f\"ur Theoretische Physik, Universit\"at Z\"urich,
Winterthurerstr. 190, 8057 Z\"urich, Switzerland}

\date{\ }
\begin{document}

\maketitle

\label{firstpage}

\begin{abstract}
We present a novel proposal strategy for the Metropolis-Hastings algorithm
designed to efficiently sample general convex polytopes in 100 or more
dimensions.  This improves upon previous sampling strategies used for free-form
reconstruction of gravitational lenses, but is general enough to be applied to
other fields.  We have written a parallel implementation within the lens
modeling framework GLASS. Testing shows that we are able to produce uniform
uncorrelated random samples which are necessary for exploring the degeneracies
inherent in lens reconstruction.
\end{abstract}


\section{Introduction}
\label{introduction}

Some inversion problems in astrophysics make it desirable to search or sample a
high dimensional solution domain $S \subset \mathbb{R}^n$ such that $S$ is bounded by
the linear constraints
\begin{equation}
\label{inequalities}
\mat Ax \le b
\end{equation}
where $\mat A \in \mathbb{R}^{m\times n}$ and $b$ is a constant vector.
A classic application is Schwarzschild's construction of  triaxial stellar
systems in equilibrium \citep{1979ApJ...232..236S}. Given a three-dimensional
discretized target density function $\rho_j$, the number of stars $c_i$ on a
given orbit $i$ is found by solving
\begin{equation}
\rho_j = \sum_i c_i \sigma_{i,j}
\end{equation}
where $\sigma_{i,j}$ is the orbit density. The orbit density is calculated
\emph{a priori} using test particles in a fixed potential corresponding to $\rho$.
However, searching the model space was not feasible at the time and only some
particular models were considered. More recent work has further developed this
technique \citep[and references therein]{1982ApJ...263..599S,1999PASP..111..129M,2006MNRAS.366.1126C}.

In this paper we consider applications to gravitational lensing. Lensing
has had quite a long history, beginning with the first direct evidence of
general relativity, but until 1979 with the discovery of the extra-solar lens
Q0957+561 \citep{1981ApJ...244..736Y} the field was largely of only theoretical
interest \citep{1964MNRAS.128..295R,1964MNRAS.128..307R}.  Today more than one
hundred strong lensing objects are known with many studied in great detail
\citep[e.g.,][]{1999AIPC..470..163K,2008ApJS..176...19F,2009ApJ...705.1099A}.
Future surveys promise to deliver thousands more.

Of utmost interest is the mass distribution of the lensing object.
Characterizing this distribution is important for understanding the properties
of galaxies and clusters \citep{2007ApJ...667..645R,2010A&A...518A..55S},
galaxy formation and evolution \citep{2010ApJ...721L...1T,2011A&A...529A..72F},
the nature of dark matter \citep{2006ApJ...648L.109C}, as well as estimating
cosmological parameters \citep{2001PhR...340..291B} and the age of the universe
\citep{2006ApJ...650L..17S,2007ApJ...660....1O,2008ApJ...679...17C}.

Crucially, the equations governing gravitational lensing are linear in the
projected mass density $\kappa$. As detailed in \secref{framework}, one can
discretize $\kappa$ onto a grid of pixels and solve for physically motivated
solutions by imposing constraints in the form of \eqnref{inequalities}.  
Several versions of this idea have been developed by
\citet{2004AJ....127.2604S}, \citet{2008ApJ...681..814C},
and \citet{2005MNRAS.363.1136K}. 

This free-form approach is more
flexible than simple analytic models, which assume a functional form of the mass
profile and may unintentionally break degeneracies. However,
this creates a large system of linear equations that is highly
underconstrained.  To understand the range of degeneracies we therefore require
a technique that can explore the space of solutions $S$.  One possible
technique is to choose a random point $x$ and accept it if $x$ lies in $S$.
This might be a reasonable method in low dimensions $n$, as is done in Monte Carlo
integration, but the probability of acceptance rapidly approaches zero as $n$
increases.  Each of the pixels in the discretization of $\kappa$ represents one
dimension and typically $n$ is greater than 100.  Complex systems, where
multiple lenses are used, can easily have more than 1000 dimensions.  The
priors can also be arbitrary, so the simplex will have a very complex shape,
although by construction it will always be convex. 

General sampling of probability distributions has been a topic of statistics
research for many years
\citep[e.g.,][]{1995...Chib...Greenberg,Robert:2005:MCS:1051451}.  In the case
of lensing, the PixeLens algorithm \citep{2004AJ....127.2604S} is frequently
used. We show, however, that the sampling of this algorithm is not
uncorrelated.  We address these details and related issues in \secref{PixeLens}
and suggest an alternative based on the Metropolis-Hastings algorithm in
\secref{new algorithm}.  In \secref{implementation} we discuss the
implementation and demonstrate in \secref{algorithm eval} that even for high
dimensions we are able to sample our solution space to achieve a uniform
uncorrelated random sample.  We also achieve significant speed improvements
over PixeLens. In \secref{outlook} we discuss future work and applications.

\section{Framework}
\label{framework}

There are two primary equations in gravitational lensing
\citep{1986ApJ...310..568B,1992grle.book.....S,2006glsw.conf....1S}. The lens
equation
\begin{equation}
\label{lens equation}
\vec\beta = \vec\theta - \nabla\psi(\vec\theta)
\end{equation}
maps an observed position $\vec\theta$ to an unobservable source position
$\vec\beta$ through the potential 
\begin{equation}
\psi(\vec\theta) = \frac1\pi\int_{\mathbb{R}^2} \kappa(\vec\theta')\ln|\vec\theta-\vec\theta'|\mathrm{d}\vec\theta'
\end{equation}
where $\kappa$ is the dimensionless projected mass density of the lensing object. The 
Fermat potential
\begin{equation}
\label{arrival time}
\tau(\vec\theta) = \frac12|\vec\theta-\vec\beta|^2 - \psi(\vec\theta)
\end{equation}
measures, up to an affine transformation, the time a photon takes to travel
from the source to the observer.  \eqnref{lens equation} corresponds to the
stationary points of \eqnref{arrival time}.  If the source varies in brightness
and the arrival times are different for different images then one can measure
the physical time delay $\Delta t_{21}$ between the light curves of
$\vec\theta_1$ and $\vec\theta_2$. 

As in \citet{2004AJ....127.2604S}, we discretize $\kappa$ into grid cells, or
pixels, centered on the lensing object and construct a system of linear
equations from $\Delta \tau \propto \Delta t$ and \eqnref{lens equation}:
\begin{equation}
\label{equalities}
\mat{C}x = d
\end{equation}
where $\mat{C} \in \mathbb{R}^{p\times k}$, $d$ is a constant vector, and $x$
is a vector containing the free parameters $\kappa$ and $\vec\beta$.  These
equations only serve to reduce the dimension of the problem $k$ by the number
of equalities $p$ since in general $p\leq k$, where $p$ is equal to twice the
number of observed images plus the number of measured time delays. This
reduction is performed with the orthogonal projection
\begin{equation}
\mathrm{proj}(x) = (\vec1 - \mat C^+ \mat C)x + \mat C^+ d
\label{projection}
\end{equation}
which takes a point $x$ to the solution set of \eqnref{equalities}.  The matrix
$\mat C^+$ is the Moore-Penrose pseudoinverse of $\mat C$.  A basis of this
affine space is given by those eigenvectors of $(\vec1 - \mat C^+ \mat C)$ with
eigenvalue equal one. We can therefore, without loss of generality, take this
reduced space of dimension $n=k-p$ to be our problem domain.

Since the space is unbounded,  we must impose constraints in the form of
\eqnref{inequalities} to limit ourselves to reasonable and physical solutions.
These constraints may derive from data, such as arrival time order or
image parity, or from Bayesian priors.  We consider only modest priors, such as the
mass must be positive everywhere, variations in $\kappa$ must be smooth, and
the local density gradient must point within $45^\circ$ of the center.  A
complete discussion can be found in \citet{2008ApJ...679...17C}. Our
choice of constraints constructs a non-empty compact solution space $S$, which
is a convex polytope, or simplex.
By our definition of the solution space, the Bayesian posterior distribution is
\begin{equation}
\label{binary prob}
P(x) \propto \left\{
    \begin{array}{rl}
    1 & \text{if } x \in S \\
    0 & \text{if } x \not\in S
    \end{array} \right.
\end{equation}
since all $x \in S$ are equally probable.  We are interested in an uncorrelated
uniform random sample drawn from $S$, which we will simply refer to as a random
sample.

\section{Revisiting the PixeLens Algorithm}
\label{PixeLens}

Earlier work used the program PixeLens \citep{2004AJ....127.2604S} to model
gravitational lenses and estimate the Hubble Time $H_0^{-1}$. The sampling
strategy employed in PixeLens is a type of random walk explained in
detail in \citet{2004AJ....127.2604S} and \citet{2008ApJ...679...17C}. Here we
summarize the algorithm and discuss some problems.

To build a set of sample points $X = \{x_1, x_2, \dots\}$ one begins
by selecting a set of vertices $V=\{v_0, v_1,\dots\}$ of $S$.  The first
sample point $x_1$ is chosen uniformly from the chord connecting $v_0,v_1$.
Each new point $x_i$ with $i>1$ is chosen randomly and uniformly from the chord
from $v_i$ through $x_{i-1}$ to the boundary of $S$.  In the limit of infinite
samples this algorithm will explore the entire simplex.  

To construct $V$, PixeLens uses the simplex algorithm
\citep{1963Danzig,Numerical-Recipes3}.  This algorithm was designed to maximize
(or equivalently minimize) a linear objective function $f(x)$ subject to linear
constraints as in \eqnref{inequalities} by moving from vertex to vertex of the
simplex in a direction that always increases $f$.  It is a standard algorithm
in the field of linear programming, where the vertex that maximizes $f$ is the
desired result.  Finding a particular vertex is not the goal of PixeLens and
so each vertex $v_i$ is found by maximizing a random objective function $f(x) = c \cdot x$
where $c = (c_1,\ldots,c_n)$ is a random vector with uniform $c_i \in [-1,1]$.

\begin{figure*}
\plotone{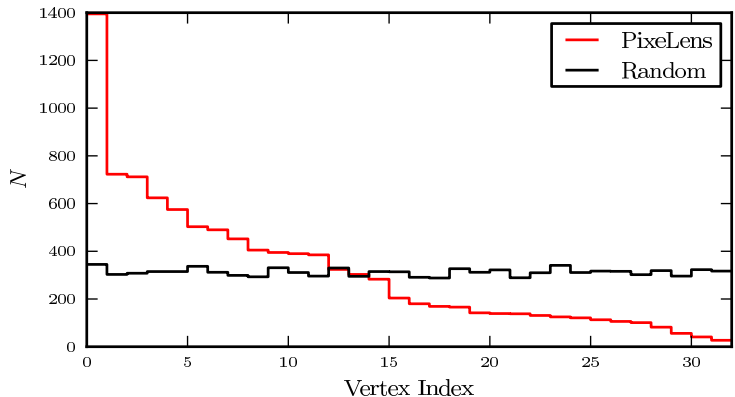}{1.2}
\caption{In the PixeLens sampling some vertices are preferred to
others when choosing a random objective function. This is shown here using a
simplex in seven dimensions with 32 vertices. The red curve is the number of
times $N$ a particular vertex was chosen using PixeLens. For comparison, a purely
random choice of the same vertices produces the black curve.}
\label{biased vertex sampling}
\end{figure*}

One issue is that randomly choosing an objective function does not randomly
choose a vertex. If the simplex is not a regular polytope there will be some
vertices that are chosen with a higher probability than others.  This is
demonstrated in \figref{biased vertex sampling} with a simplex in seven
dimensions with 32 vertices. The vertices have been enumerated and sorted by the
number of times they were chosen. Clearly some vertices are highly preferred.
Even in high dimensions where the choice of a particular vertex is unlikely to
occur again, vertices that are particularly acute will be more likely.  By not
choosing vertices at random the algorithm prefers some regions over others
which leads to correlations in the final sample and not all directions will be
adequately explored. Vertex selection is also not invariant under some general
coordinate transformation $x \rightarrow \mat T x$ for an invertible
matrix $\mat T \in \mathbb{R}^{n\times n}$.  Even a change in units may affect
the sample distribution and therefore the inferred physical parameters.

Correctly choosing vertices at random is itself a difficult problem.  There
exist several methods to enumerate the vertices 
\citep[e.g.,][]{springerlink:10.1007/BF02293050, 1983jstor...dyer} but unfortunately the
number of vertices has a huge combinatorial upper bound of
\begin{equation}
{m + \lfloor\frac12 n \rfloor \choose m}
+
{m + \lceil\frac12 n \rceil - 1 \choose m}
\end{equation}
where $m$ is the number of inequalities \citep{nla.cat-vn2686816}.

To avoid the enumeration we modified the simplex algorithm to randomly walk
between the vertices. This dispenses with the objective function and simply
selects a neighboring vertex to which to move. While this does improve the
sampling of the vertices (see \figref{biased vertex sampling}), it is not
without its own problems.  If a vertex has many close neighbors, as is likely
in high dimensions, the random walk will tend to stay in one region before
moving large distances.  To compensate, one must run for a long time. The
process of moving to a new vertex is computationally costly, however, and
incurs numerical error that quickly dominates after too many iterations.

\begin{figure*}
\plotone{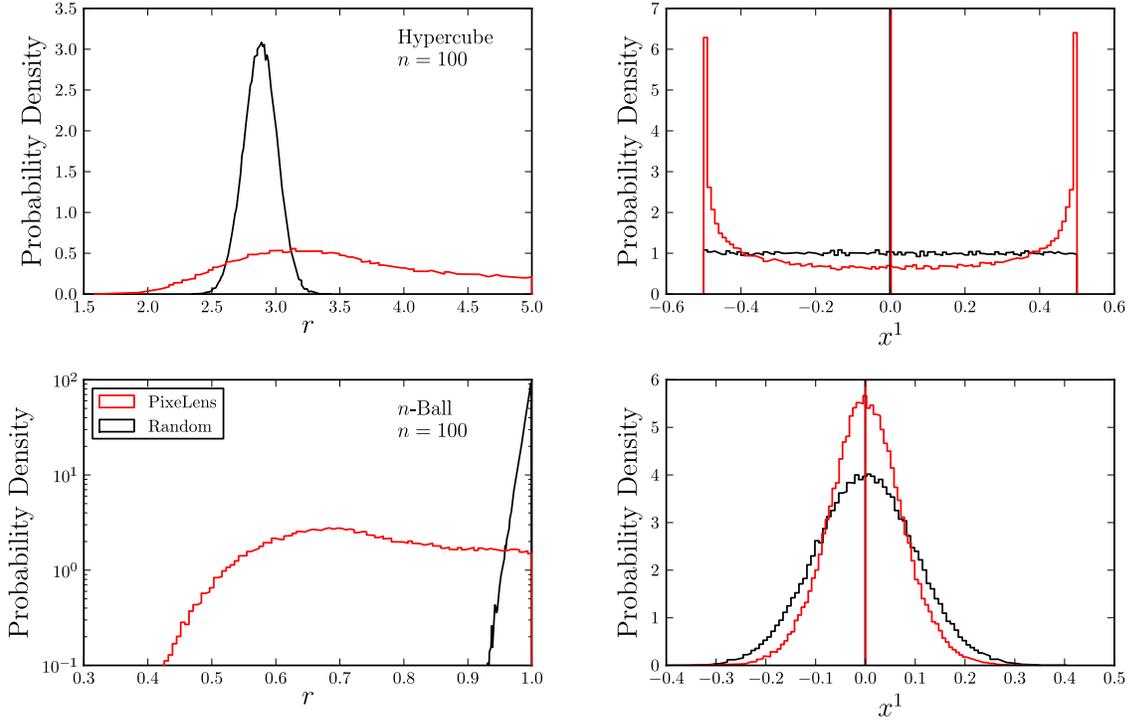}{1.8}
\caption{A comparison of sample sets $X_P$ and $X_R$ for the PixeLens sampling algorithm \textbf{(red)} and
random sampling \textbf{(black)}, respectively. Each set has 100,000 items.  The means of
the plotted distributions (vertical lines) for both the hypercube and $n$-ball
samples are nearly identical, whereas the deviations are not.
\textbf{(Top left)}
We take the sample space $S$ to be a unit cube centered at the origin in 100
dimensions and show the probability distribution function of $r=|x|$ for all $x
\in X$.
\textbf{(Top Right)}
The PDF for the single coordinate $x^1$ of each sample point. Marginalizing over the other
coordinates we expect the probability to be uniformly
one. Points from $X_P$ clump along the chords connecting vertices.
\textbf{(Bottom left)} 
Similarly, we plot the PDF where $S$ is a $n$-ball.
\textbf{(Bottom right)}
Marginalizing out a single coordinate we see that a random sample has a broader
distribution.}
\label{biased interior sampling}
\end{figure*}

Another issue is that the PixeLens algorithm is based on the vertices of $S$.
The algorithm produces samples that do not follow the target probability
distribution function $P(x)$, even if the vertices are chosen randomly.  In
\figref{biased interior sampling} we demonstrate this for a 100 dimensional
hypercube and $n$-ball, where the vertices have been randomly selected \emph{a
priori}  to avoid the PixeLens vertex selection algorithm. In the case of the
$n$-ball we have chosen a random set of points on the surface to be the
vertices. Points chosen from the hypercube tend to lie in the corners, while
points from the $n$-ball are more closely clustered in the center.  In general,
the points tend to clump along the chords connecting vertices.  In both cases
the PixeLens sampling is markedly different than a random sample, although the
means are nearly identical.  

The sample should be uniformly distributed in $S$ in order to be able to
perform statistical analysis on it.  For this reason, we chose to explore
an alternative method based on a random walk that does not depend on the
vertices of the simplex.

\section{A New MCMC Proposal Density}
\label{new algorithm}

The Metropolis-Hastings algorithm \citep{1953JChPh..21.1087M,1970...Hastings} is a well known method
to sample the probability distribution $P(x)$ by generating a Markov chain $X =
\{x_1, x_2, \dots\}$.  A sample $x'$ is selected from a proposal density
function $Q(x', x_i)$ given the current sample $x_i$ and if
\begin{equation}
\alpha < \frac{P(x')Q(x_i, x')}{P(x_i)Q(x',x_i)}
\end{equation}
where $0 \le \alpha \le 1$ is chosen from a uniform distribution, then $x'$ is
accepted and $x_{i+1} = x'$. If $x'$ is rejected the current point is duplicated as 
$x_{i+1} = x_i$. We will assume that $Q$ is symmetric, i.e., $Q(x',x_i) =
Q(x_i,x')$, so that the chance of moving from $x_i$ to $x_{i+1}$ is the same as
moving from $x_{i+1}$ to $x_i$.

One possibility for $Q$ is to simply move by an arbitrary amount in a random
direction but the chain may become trapped in narrow regions, especially in
high dimensions.  To account for the shape of $S$, $Q$ is often taken to be a
multivariate Gaussian distribution $\mathcal{N}(x_i,\Scovar)$, where $\Scovar$
is the covariance matrix of the sample $X$.  This matrix is an estimate of the
covariance matrix $\mat\Sigma$ of $S$ and can be progressively calculated as
the chain is built.  Such adaptive chains are no longer Markovian because the
reversibility is broken, but one can run an initial adaptive burn-in phase
before beginning the Markovian chain with $\Scovar$ fixed at its last value.

Selecting $x'$ from $\mathcal{N}(x_i,\Scovar)$ is equivalent to selecting  $y'$
from the distribution $\mathcal{N}(0,\mat{1})$%
\footnote{
The probability density function for $y'$ is $\displaystyle f(y') = (2\pi)^{-\frac n2}e^{-\frac12\Vert
y'\Vert^2}$.}
and setting $x'=x_i+\mat E\,\mat\Lambda^{1/2}y'$, where $\mat E$ is the matrix
of the eigenvectors $e_j$ of $\Scovar$ and $\mat\Lambda$ is the diagonal matrix
with the corresponding eigenvalues $\lambda_j$. In other words, we move along a
randomly selected direction accounting for the shape of $S$ through the
eigenvalues.

\begin{figure*}
\plotone{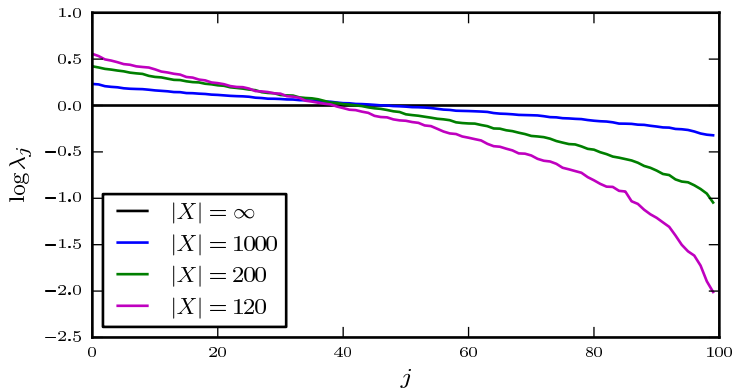}{1.2}
\caption{The sorted eigenvalues of the sample covariance matrix of three sets
of random points in a 100 dimensional cube with side length $\sqrt{12}$. The set
size is $|X|$. With infinite samples we expect all eigenvalues to be equal to
unity, but due to the high dimensionality, a random sample does not fully
estimate each direction, especially for low $|X|$.}
\label{poor eigenvalues}
\end{figure*}

For a reasonable estimate of $\mat\Sigma$, particularly in high dimensional
spaces, it is important to have $\gg n$ points.  When only $\gtrsim n$ points
are known, some of the eigenvalues of $\mat\Sigma$ are strongly underestimated
simply due to poor sampling of the space.  Even if true random samples were to
be drawn directly from $S$, the shape would be incorrectly estimated.  In
\figref{poor eigenvalues} all the eigenvalues of a 100 dimensional cube should
be equal, but for small sample sizes $|X|$ this is clearly not the case.
Poorly estimated eigenvalues cause the standard proposal density to undersample
$S$ in the direction of the corresponding eigenvectors, even as new points are
added to the chain; the new points reinforce the bias that was present in the
original $\Scovar$. 

The key improvement from this paper is to use the constraint information of
\eqnref{inequalities} to hint at the shape of $S$ and achieve a reasonable
proposal density despite having a small sample size.  We do this in the
following way. Let $X = \{x_1, x_2, \dots, x_k\}$ with $k \geq n+1$ be a set of
points in $S \subset \mathbb{R}^n$.  These points may be chosen in any fashion,
but the sample covariance matrix $\Scovar$ must be invertible, i.e., all
$x_i$'s do not lie on the same hyperplane of $\mathbb{R}^n$, and thus the set
of the eigenvectors of $\Scovar$ is an orthonormal basis of $\mathbb{R}^n$.
Since $S$ is convex, the mean $\mean X$ will also be in $S$.  Extending the
eigenvector $e_j$ from $\mean X$ intersects the boundary of $S$ at two points:
one in the positive and one in the negative direction.  The distance $d_j$
between each pair of boundary points is taken as an estimate of the size of $S$
along $e_j$.

\begin{figure*}
\plotone{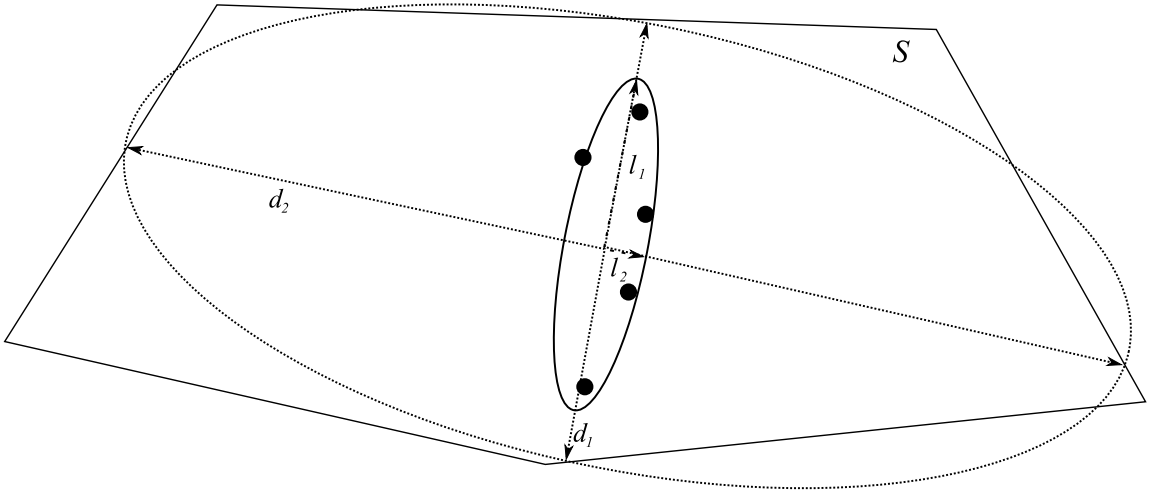}{1.8}
\caption{Our modification to the choice of proposal density $Q$ is show
schematically. The eigenvectors and corresponding eigenvalues
$\lambda_1=l^2_1,\lambda_2=l^2_2$ of the covariance matrix of an initial set of points do
not accurately capture the shape of the simple $S$. This is a problem only for
high dimensions. We show a two dimensional drawing for simplicity.  Since the
boundaries of $S$ are known, we extend the eigenvectors until we reach the
boundary.  We substitute the eigenvalues with $d_1^2/12,d_2^2/12$.}
\label{algorithm}
\end{figure*}

Our modification takes $Q$ to be the multivariate Gaussian distribution
$\mathcal{N}(x_i, \Scovarmod)$, where $\Scovarmod = \mat E\mat D\mat E^T$ and
$\mat D$ is the diagonal matrix of the new $\sigma^2_j = d^2_j / 12$.
We therefore select $x'=x_i+\mat E\,\mat D^{1/2}y'$.  The ellipsoidal shape of
$Q$ is thus adjusted by substituting $\mat\Lambda$ with $\mat D$ to better
approximate the shape of $S$.  In \figref{algorithm} we show schematically the
modification. The initial set of points inadequately samples the horizontal
direction and therefore the eigenvalues $\lambda_1,\lambda_2$ of the covariance
matrix are small in that direction. The new distances $d_1,d_2$ are a much
better approximation and are taken in place of the eigenvalues.  This strategy
encourages the movement along directions that have been poorly sampled and
therefore have a very small variance. This is an important distinction to
so-called Hit-and-Run algorithms \citep{HitAndRun1993} that step in a random
direction and may require many iterations to move through narrow spaces.

Metropolis algorithms in high dimensions can often be inefficient.  If the
average step length is too big, almost all proposals will fall into low
probability regions and be rejected, whereas if the step length is too small,
almost all proposals will be accepted but sampling the space will be very slow.
The optimal is somewhere in between and can be reached if we regulate the step
length by multiplying the $\sigma_i$'s with some scaling constant $c$ such that
the acceptance rate is roughly $25\%$ \citep{Gelman96}. In our case $P(x)$ is
not a multivariate Gaussian distribution and its shape varies from case to case
and thus there is no single $c$. However, we found that for $c \sim 1/n$ the
acceptance rate remains reasonable.

Assuming $c=4/n$, the random walk will typically translate to an average
distance of $t_1=4\left(\sum_j \sigma_j^2/n^2\right)^{1/2}$ after one
step and $t_s=t_1\sqrt{s}/2$ after $s$ steps assuming an acceptance rate of
$25\%$. In order to produce two uncorrelated points in $S$, we must make $s =
N_s$ steps such that the average traveled distance is of the order of the
simplex diameter $D$.  In the case of the hyperrectangle or similar shaped
simplices
\begin{equation} 
\label{simplex diameter} 
D \sim 2\sqrt{\sum_j \sigma_j^2} = \frac{n}{2} t_1=t_{n^2} 
\end{equation} 
and therefore $N_s = O(n^2)$ steps are needed. In other cases, such as a regular
$n$-simplex, where the approximation for $D$ in \eqnref{simplex diameter} does
not hold, $N_s > O(n^2)$ steps may be required. This is not a typical scenario
however in lensing, as we demonstrate in \secref{algorithm eval}.

The strength of this algorithm is that it is not sensitive to the dispersion of
the starting set of points $X$ but only to its mean $\mean X$.  When
$\mean X$ is not a good estimate of the center of $S$, the algorithm
will need some time to remove the starting bias. 

\section{Implementation}
\label{implementation}

We have implemented our modified sample strategy in a new gravitational lens
modeling framework called GLASS. This framework is specifically designed for
free-form lens modeling and to allow for easy modification of modeling
strategies and priors. Furthermore, we are able to immediately test the
implementation by comparing with lensing theory and published results.

As discussed in the previous section, the proposal density $Q$ depends on an
estimate for the size of the simplex $S$. We estimate the size by measuring the
distances $d_j$ from the current sample chain mean $\mean X$ to the boundary
of $S$ following the estimated eigenvectors $e_j$.
These diameters are best estimated if
$\mean X$ is close to the simplex mean $\mean S$ and the vectors $e_j$ are
aligned with the true eigenvectors of $S$. 

As is often done \citep{Numerical-Recipes3}, the Markov chain is restricted to
move coordinate-wise along the eigenvectors $e_j$, rather than in a random
direction.  We rotate $S$ such that these eigenvectors coincide with the standard
basis. This provides a significant performance improvement since only one
coordinate needs to be updated. In addition, the constraints
\eqnref{inequalities}, typically numbering a few times $n$ in lensing, must
only be checked in one coordinate.  After $N_s$ steps the last point is rotated
back into the original coordinate system and appended to $X$. With $N_s=O(n^2)$
the running time to produce one sample is $O(n^3)$. While PixeLens also has a
theoretical running time of $O(n^3)$ our implementation has a reduced scaling
constant resulting in significant performance gains.

Our implementation begins by finding the point $x_0$  where the temporary
variable $t$ is maximized subject to $\mat A x_0 + t \leq b$.  For this we use
the simplex algorithm but any linear programming algorithm will suffice.  The
point $x_0$ is inside $S$ and in some sense ``far'' from the boundaries. 

Initially, the chain walks along the eigenvectors of the matrix $\mat1 - \mat
C^+\mat C$.  The chain is run for a burn-in phase where we collect $N_b$
samples.  We typically let $N_b=10n$.  After the first $2n$ samples, and
subsequently after each $N_b/10$ samples, we updated $Q$ by calculating
$\Scovarmod$ and then continue walking along the respective eigenvectors. The
scaling constant $c$ is adjusted to hold the acceptance rate around 25\%.

After the burn-in phase, we fix $\Scovarmod$ and $c$ at their final values and run a
new chain for as many samples as are desired. In both phases we can run several
Markov chains in parallel as long as we ensure that all threads use the same
$\Scovarmod$. We have tested this on a shared memory machine using up to 48 CPUs.

As we move to higher dimensional spaces we expect that accumulation of
round-off error will tend to produce departures from the equality constraints
in \eqnref{equalities}.  To remove this numerical error we ensure that a sample
point $x$ lies on the simplex by using \eqnref{projection} to project it back
onto the simplex.

\section{Sampling Evaluation}
\label{algorithm eval}

The stationary distribution of the sample set $X$ from any general MCMC strategy
will be the target probability distribution $P(x)$. This is only reached though
for a sample size much greater than $n$. In practice this is not feasible in
high dimensions and we must limit our sample size to $\gtrsim n$.  As we
demonstrated in \figref{poor eigenvalues} the eigenvectors of a small random
sample will not be able to fully describe the solution space but for lensing
statistics this is sufficient as we typically marginalize over many parameters.
As we want to obtain a uniform uncorrelated random sample, we compare our MCMC
implementation to a random sample from a hyperrectangle and a regular
$n$-simplex in 100 dimensions, while varying $N_s$ to measure convergence. 

\begin{figure*}
\plotone{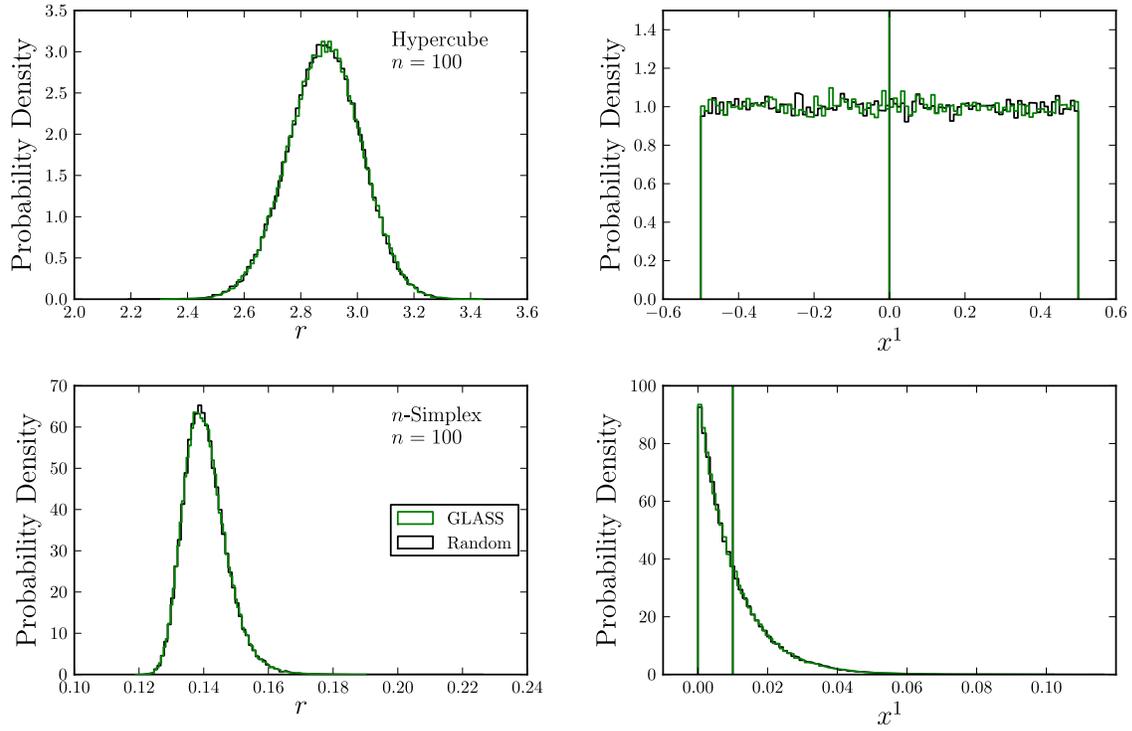}{1.8}
\caption{Repeating the tests shown in \figref{biased interior sampling} using
GLASS. With the MCMC implementation we correctly recover the same distributions
as a random sample. 
\textbf{(Left plots)} The distance $r=|x|$ of each point from the origin. 
\textbf{(Right plots)} The distribution of the first coordinate of $x \in X$, marginalizing over
the others.  The $n$-Simplex lies on only one side of the origin. The vertical
lines mark the mean of each sample.
}
\label{unbiased interior sampling} 
\end{figure*}

The hyperrectangle 
\begin{equation}
H_n = \{(x_1,x_2,\dots,x_n) \in \mathbb{R}^n\;|\; 0 \leq x_j \leq 1/j\}
\end{equation}
is straightforward to sample directly and the $n$-simplex 
\begin{equation}
S_n = \{(x_1,x_2,\dots,x_{n+1}) \in \mathbb{R}^{n+1} \;|\; 0 \leq x_j, \sum_j x_j = 1\}
\end{equation}
is only slightly more involved%
\footnote{To generate a uniform random point in the regular $n$-simplex $S_n$ choose
an $(n+1)$-vector $V$ of i.i.d.~numbers drawn from an exponential
distribution.  Then $V' = V/\sum_j V_j$ is a point in $S_n$ \citep{devroye:1986}.}. We expect the
simplex of a real lens system to be similar to $H_n$ since it has $2n$ inequalities
and $2^n$ vertices. In addition, we also test $S_n$ as an extreme example.  We
repeated the tests of PixeLens from \secref{PixeLens} using the hypercube and
$S_n$ with $N_s=n^2$ and $N_s=n^{2.5}$, respectively, and show the results in \figref{unbiased interior sampling}. We are able
to match the expected distributions of a random sample perfectly.

\begin{figure*}
\plotone{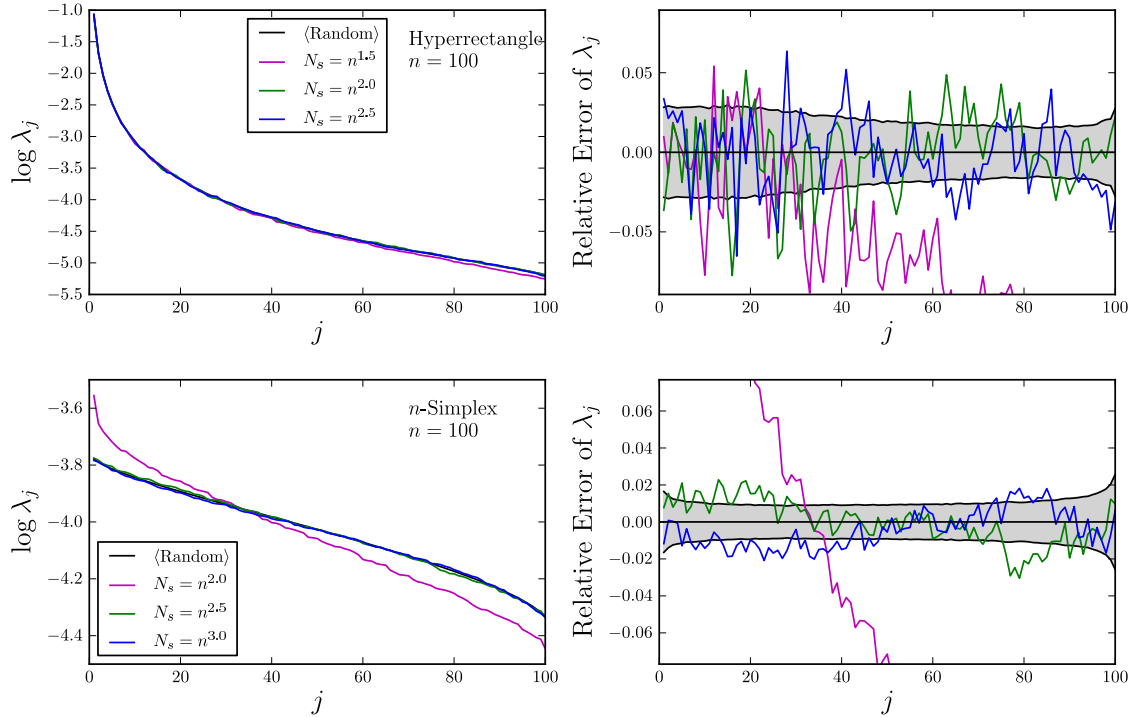}{1.8}
\caption{A convergence test of the MCMC sample eigenvalues $\lambda_j$ with
increasing $N_s$. For comparison, several hundred random sets were
generated.  The average of the random sets is shown in the left
plots. The right plots normalize the MCMC eigenvalues to the mean of the random
values. The grey band shows the $1\sigma$ deviations of the random sets.
The hyperrectangle \textbf{(top)} converges by $N_s\sim n^2$ as
predicted from \eqnref{simplex diameter}, but the $n$-Simplex \textbf{(bottom)}
with its different shape requires a larger $N_s \sim n^{2.5}$.}
\label{eigenvalues}
\end{figure*}

We test the global properties of our samples $X_H, X_S$ against the random
samples $R_H,R_S$ by comparing the eigenvalues of the respective sample
covariance matrices.  Each sample set contains 1000 points. In
\figref{eigenvalues} we show the sorted eigenvalues for these samples.  As
expected from \eqnref{simplex diameter} the hyperrectangle converges at $N_s
\sim n^2$. The $n$-simplex requires a larger $N_s$, as mentioned in \secref{new
algorithm}, because the estimate of the diameter from \eqnref{simplex diameter}
is no longer valid.  In the right panels, we have taken several hundred random
sample sets and plotted the $1\sigma$ deviations. The plots are normalized to
the mean of these random sets.  The volume of a simplex can be well
approximated by $(\det \Scovar)^{1/2} = \prod_j \sqrt{\lambda_j}$.
\tabref{simplex volumes} shows the volumes of all the computed samples.  The
uncertainties have been calculated from 1000 independent runs. Our sampling
strategy is in excellent agreement with random samples. 

\begin{figure*}
\plotone{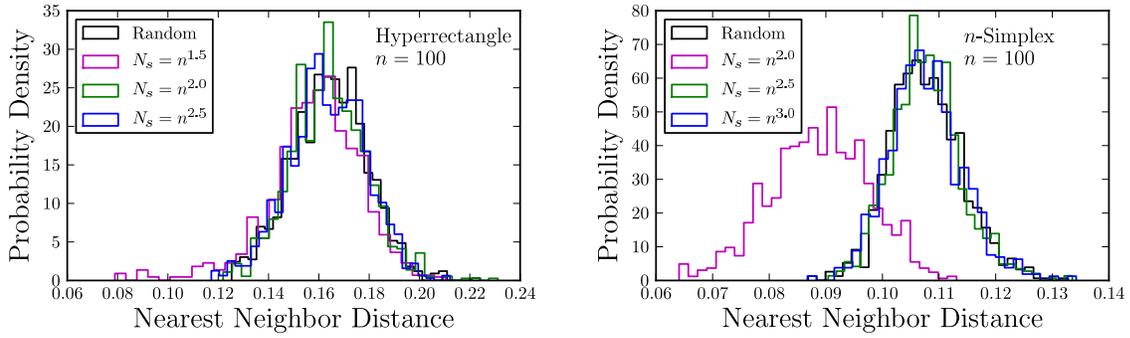}{1.8}
\caption{The probability that the nearest neighbor of a sample point is at a given
distance in the hyperrectangle \textbf{(left)} and $n$-simplex \textbf{(right)}.
The histograms produced by GLASS show similar convergence as in \figref{eigenvalues}.}
\label{nearest distance}
\end{figure*}

\begin{figure*}
\plotone{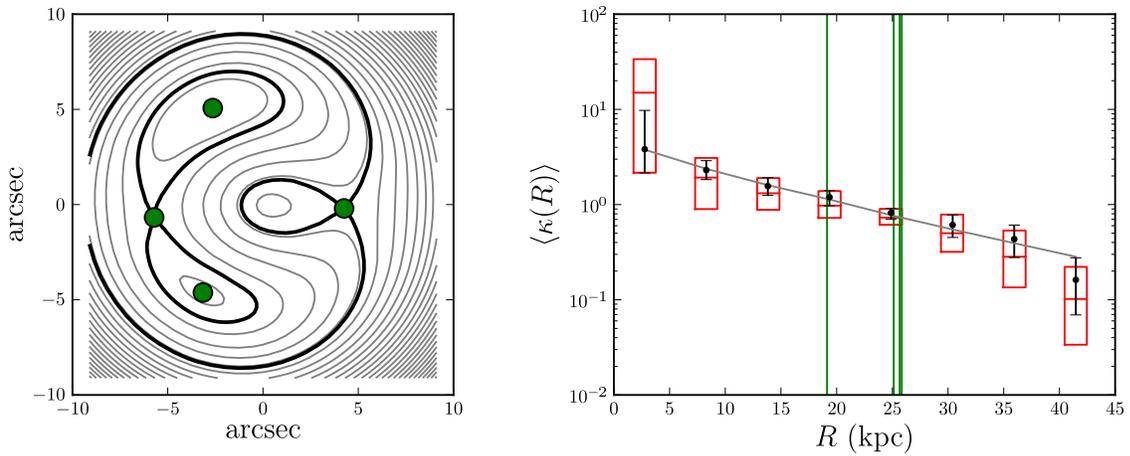}{1.8}
\caption{The reconstructed lens model of a source producing four images.
\textbf{(Left)} The arrival time surface of the lens. The images
\textbf{(dots)} are at the stationary points of the surface. \textbf{(Right)}
The radial density profile of the original lens mass \textbf{(grey)} compared
to the ensemble of possible models produced from the MCMC sampling
\textbf{(black error bars)}. The boxes \textbf{(red)} show the
results obtained from PixeLens for the same lens where the error estimates
clearly favor shallower models suggesting that the old algorithm over-samples
some regions of the parameter space. The radial image positions are marked by
vertical lines.}
\label{lens recon}
\end{figure*}

\begin{figure*}
\plotone{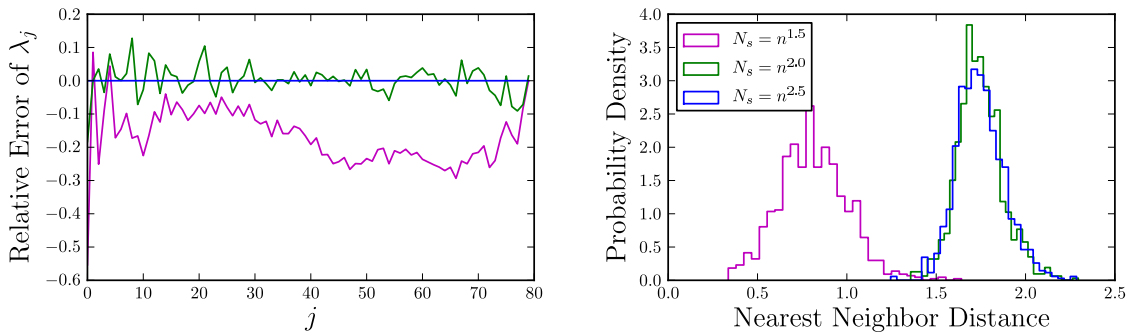}{1.8}
\caption{A convergence test of the sample eigenvalues \textbf{(left)} and the
probability of nearest neighbor distances \textbf{(right)} using the realistic
lens model depicted in \figref{lens recon}. The number of steps $N_s$ used in
the MCMC walk is increased until the histograms of nearest neighbor distances
converge.  An accept/reject random sample is impossible to obtain given the
unknown shape of the simplex.  The left plot has been normalized to the
$N_s=n^{2.5}$ case. For $N_s \gtrsim n^2$ the eigenvalues are stable.}
\label{lens ev}
\end{figure*}

\begin{table*}
\begin{minipage}{130mm}
  \caption{Estimated volumes for the tested simplices of $H_n$, $S_n$, and the
  gravitational lens for various values of $N_s$, demonstrating convergence. The
  volumes are calculated from the eigenvalues shown in \figref{eigenvalues} and
  \figref{lens ev}. The column labeled random is calculated via direct random
  sampling, which is not possible for the lens. Where the value is red, the
  volume is underestimated compared with the random sample.}
  \begin{center}
    \begin{tabular}{c|llllll}
    \hline
     & $n^{1.5}$ & $n^{2}$ & $n^{2.5}$ & $n^{3}$ & Random \\
    \hline
    $H_n$ & \darkred{$0.244\pm0.044$} & $9.1\pm1.5$ & $8.9\pm1.4$ & --- & $9.0\pm1.4$ & \e{-214}\\
    $S_n$ & ---  & \darkred{$0.094\pm0.043$}  & $1.64\pm0.51$  & $1.62\pm0.46$ & $1.63\pm0.52$ & \e{-202}\\
    Lens & \darkred{$(9.5\pm8.5)\e{-8}$} & $0.94\pm0.19$ & $1.21\pm0.24$ & --- & --- & \e{-93}\\
    \hline
    \label{simplex volumes}
    \end{tabular}
  \end{center}
\end{minipage}
\end{table*}

While the eigenvalues paint a global picture, we also tested the local
properties of our sample. In particular, we looked at the distribution of
nearest neighbor distances.  In \figref{nearest distance} we show this
distribution for $H_n$ and $S_n$ and for increasing $N_s$.  A misalignment with
a random sample or multiple peaks are indicators of a correlated sample, such
as clumped points.  The chains with too low $N_s$ are unable to traverse across
the simplex resulting in sample points which are too close to each other.  By
the time the eigenvalues converge the local distribution also converges.

Finally, we tested our implementation with a simulated triaxial lens mass. The
four image positions and respective time delays were calculated using a root
finding algorithm built into GLASS.  We supplied the value of $H_0$, all the
time delays, and all the image positions to the algorithm without error bars to
test the effectiveness of the method. The problem, however, still remains
heavily underconstrained. In the near future, we will explore in detail the
effects of relaxing these assumptions in a variety of different lens systems.

For the reconstruction, we used a grid of 225 pixels but we assumed radial
symmetry to reduce the number of independent pixels to 113.  Together with the
unknown source position the problem lies on a 104 dimensional affine space.
The reconstructed average arrival time surface and images are shown in the left
panel of \figref{lens recon}.  The right panel compares the inferred surface
density profile with uncertainties (black error bars) to the original lens
profile.  The constraint information on the mass profile is contained within
the image positions (vertical lines) and therefore the pixels outside
are not expected to be well fit.  In general, though, this reconstruction is
excellent where the information content is highest. 
As discussed in \secref{introduction}, by sampling the solution space, we are able
to explore the degeneracies which simple models cannot.  For comparison, we
also show the results for the same lens obtained using PixeLens (red boxes).
Although the results are similar, the PixeLens error estimates favor shallower
or nearly flat models, again suggesting that the old algorithm over-samples
some regions of the parameter space as shown in the upper-right panel of
\figref{biased interior sampling}.

We also performed the same eigenvalue and nearest neighbor analysis as before
but because we are unable to directly sample the solution space we can only
change $N_s$ until we converge.  As expected and shown in \figref{lens ev} we
converge when $N_s \sim n^2$. 

\section{Outlook}
\label{outlook}

Our novel proposal strategy for the Metropolis-Hastings algorithm allows
sampling of general convex polytopes in 100 or more dimensions.  We have
implemented an efficient parallel version of the algorithm in the gravitational
lens modeling framework GLASS so that we may apply the strategy to large
lensing problems exceeding 1000 dimensions. GLASS will be publicly available
for download in the near future.

Several future applications are possible.  Multiple redshift sources carry
information of the cosmological distances, which in turn depend on the
cosmological parameters.  Considering the statistical dispersion of the
parameter space, one could in principle be able to infer the cosmological
parameters in a Bayesian framework.  In order to achieve this, a uniform sample
of the solution space is needed.  

Previous work on estimating the Hubble Time has used systems of up to eighteen
lenses \citep{2010ApJ...712.1378P}. New lenses can be included to further
constrain this value but each additional lens increases the dimensionality of
the space by $\sim100$ making this current work essential for such upcoming
projects.

\section*{Acknowledgements}

The authors would like to thank Prasenjit Saha for careful reading of the
manuscript and extremely helpful advice.

\bibliographystyle{mn2e}

\bibliography{ms}

\bsp

\label{lastpage}

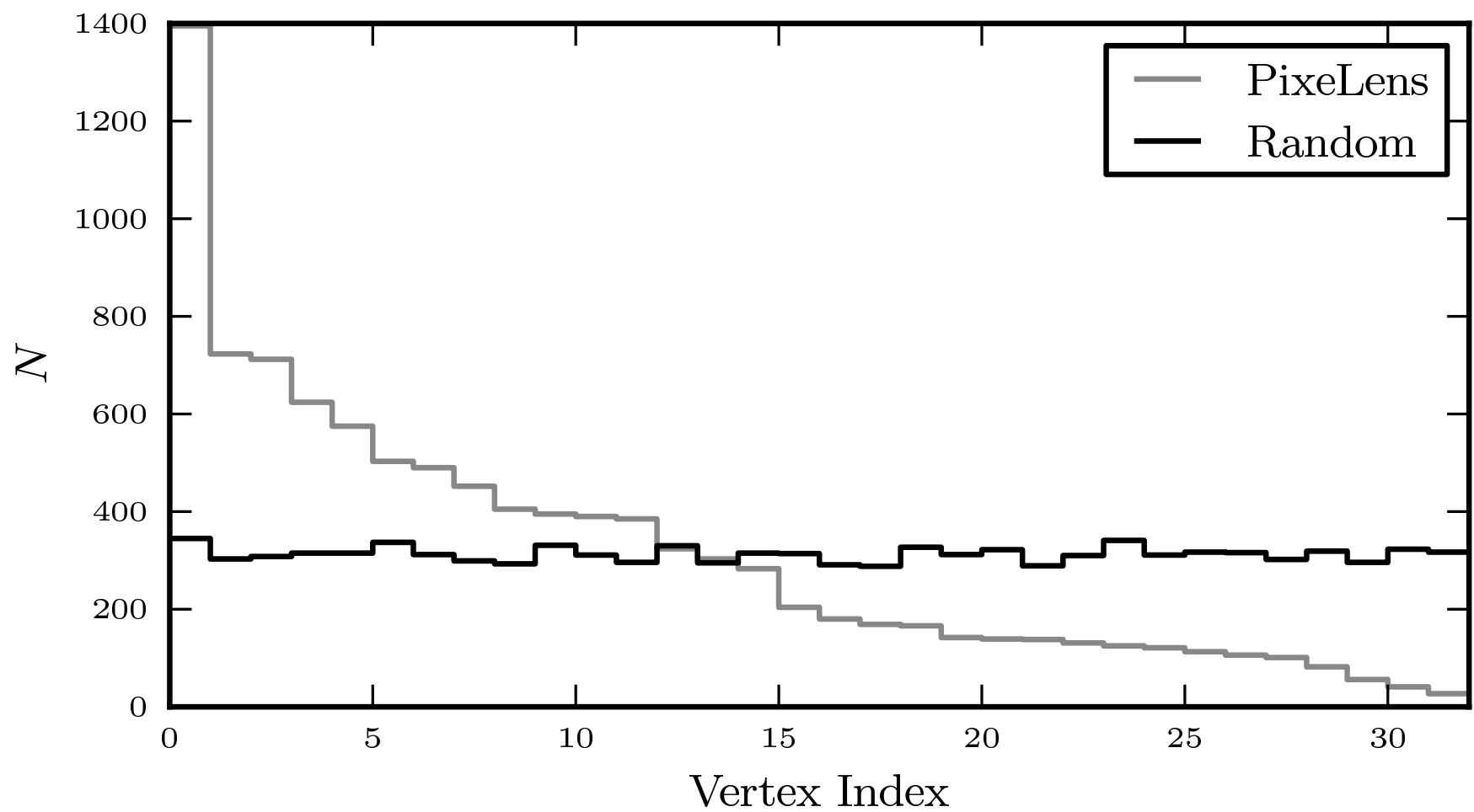
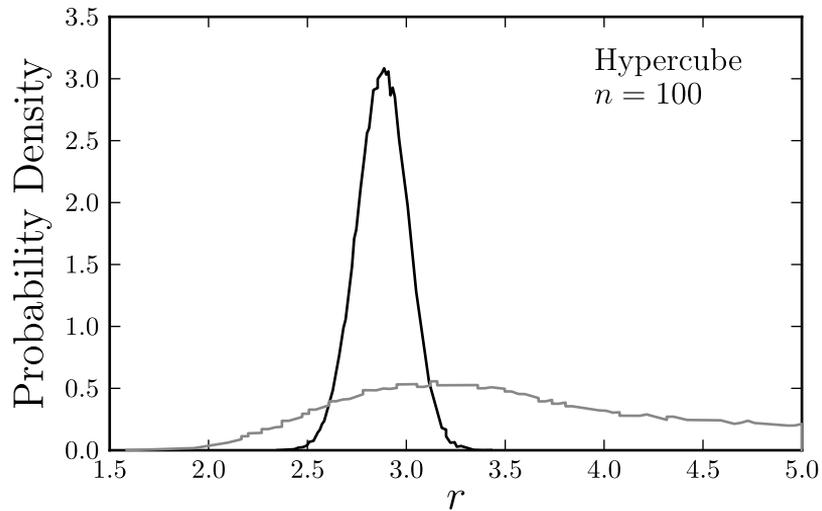
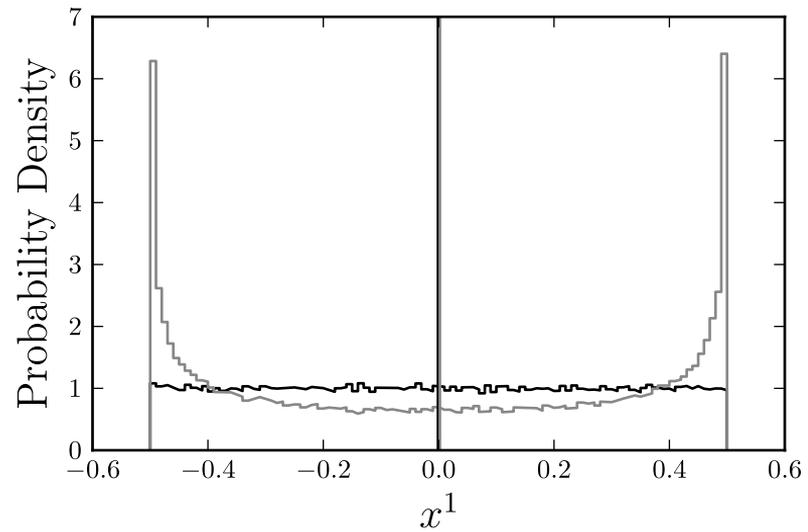
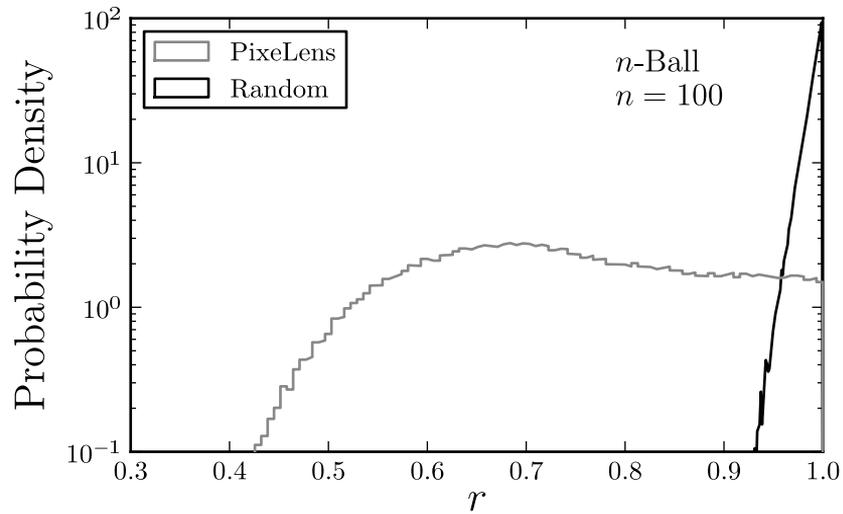
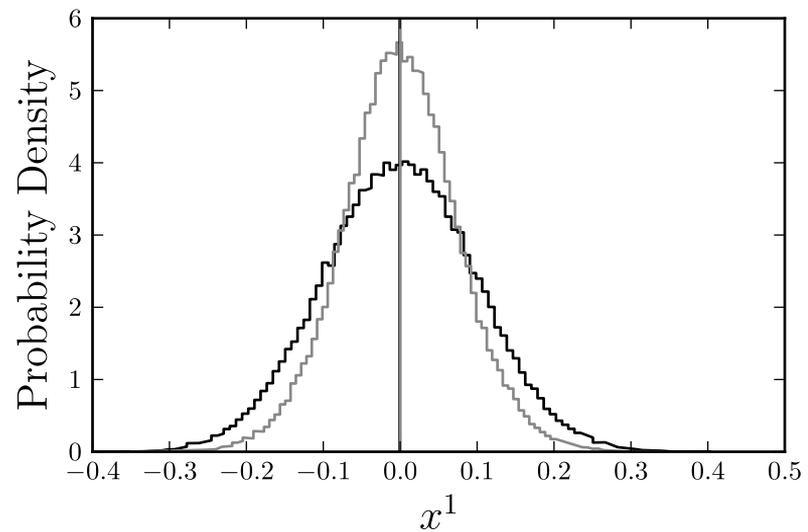
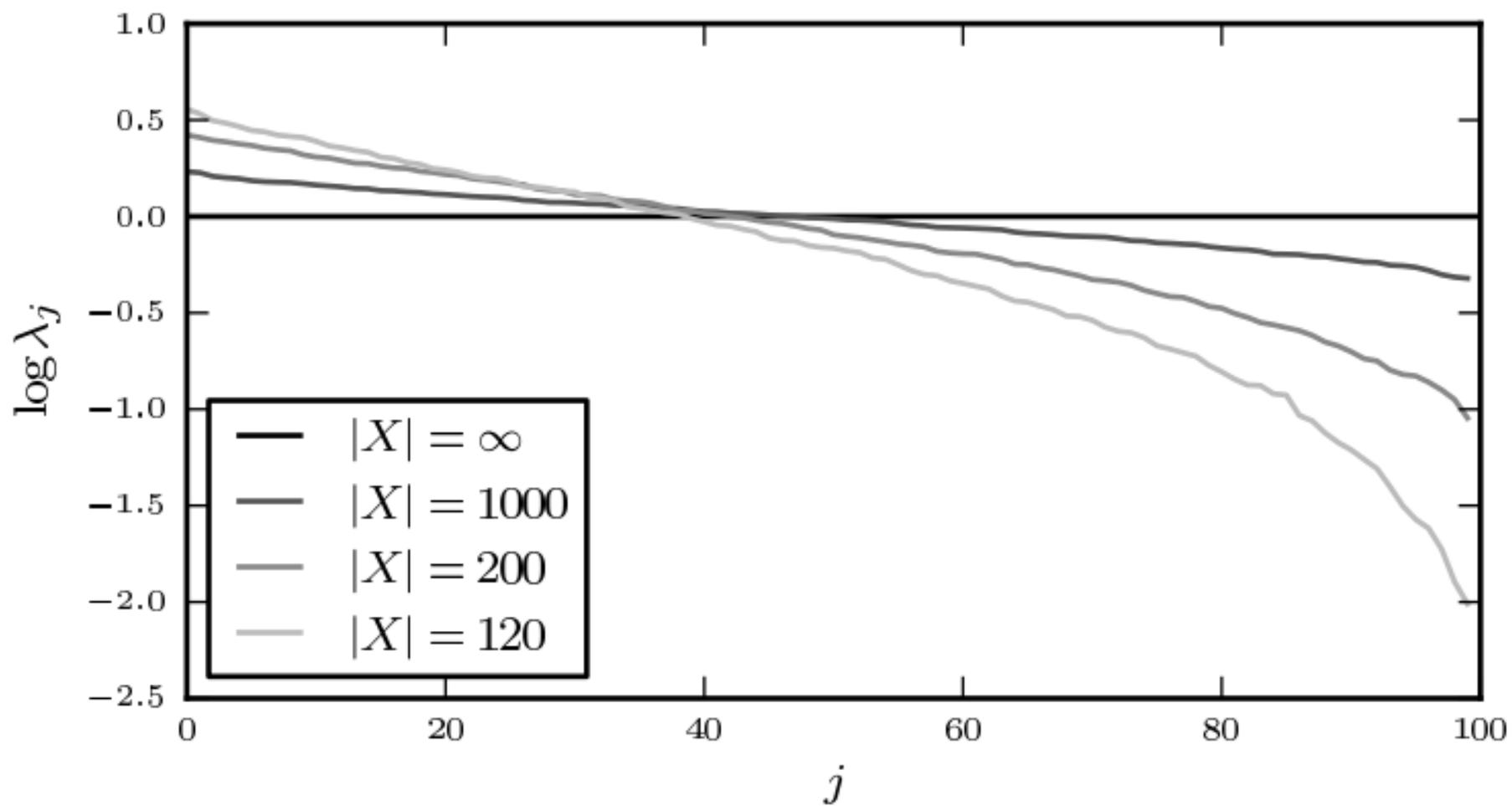
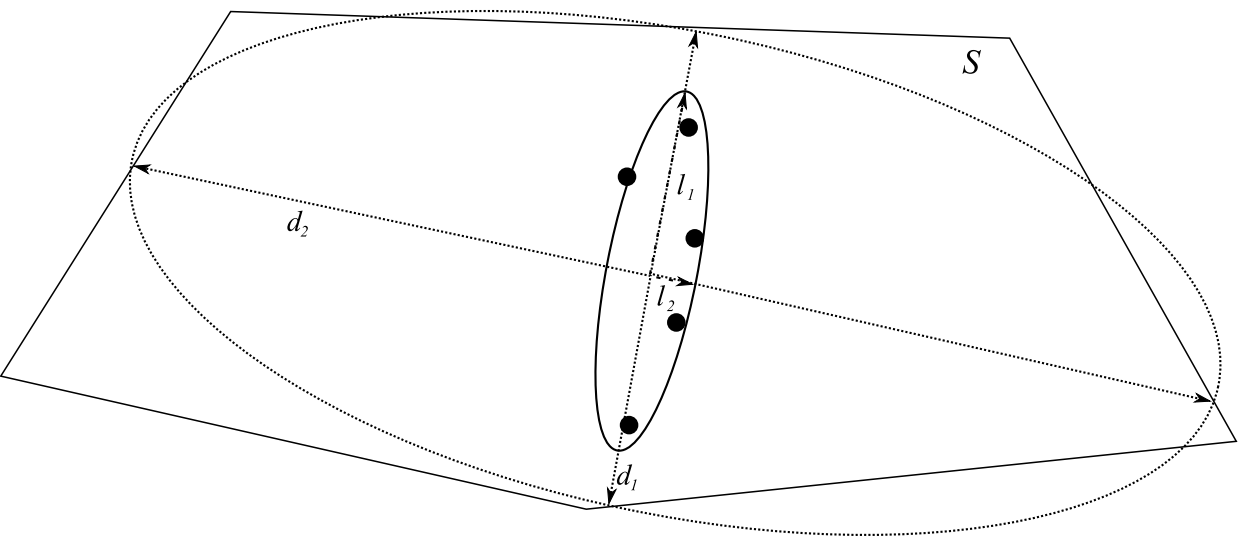
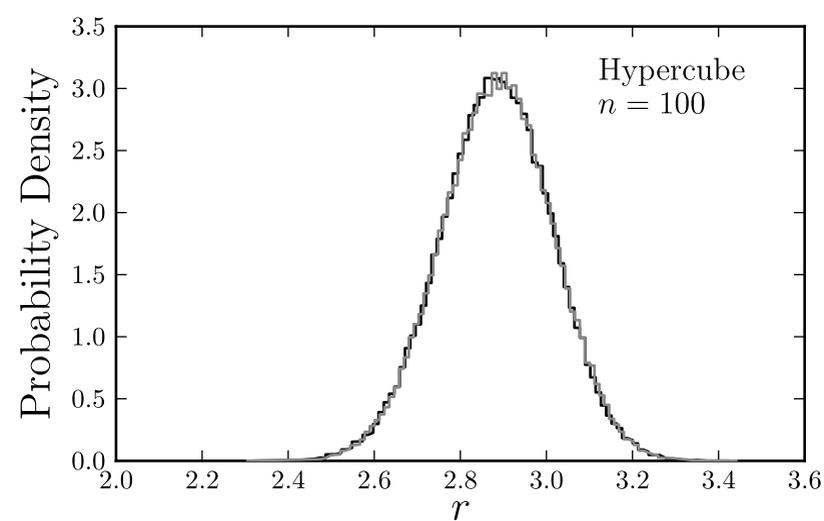
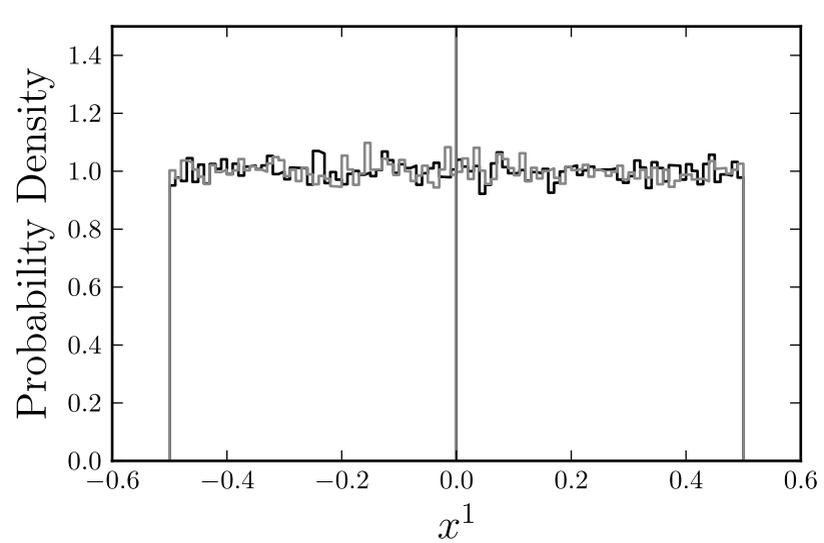
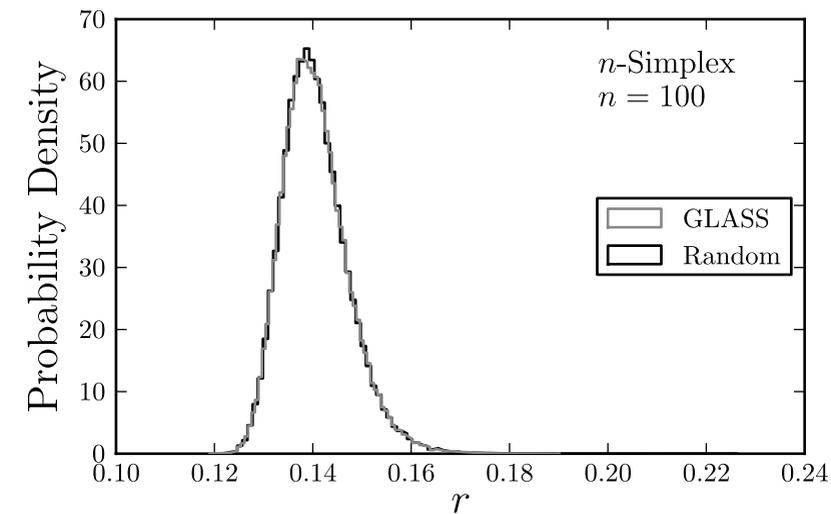
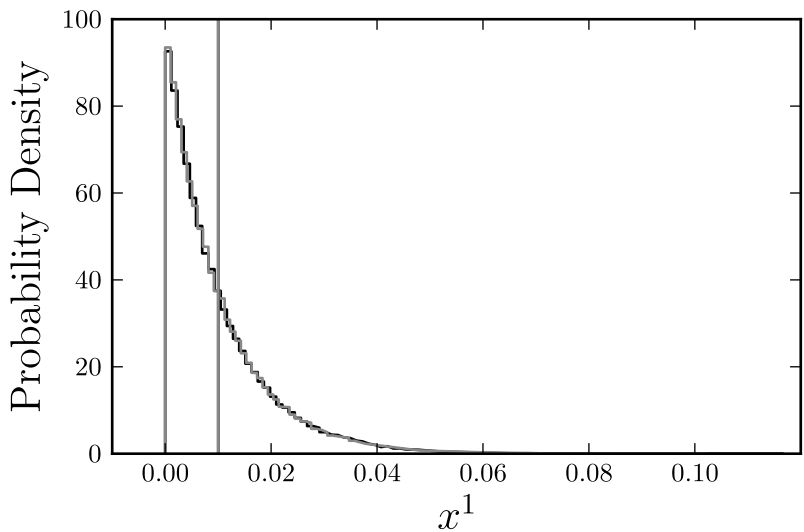
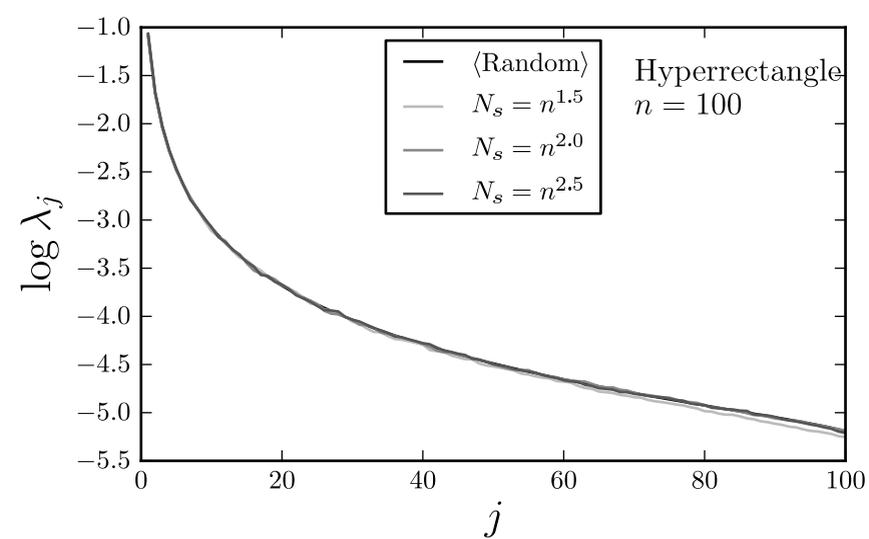
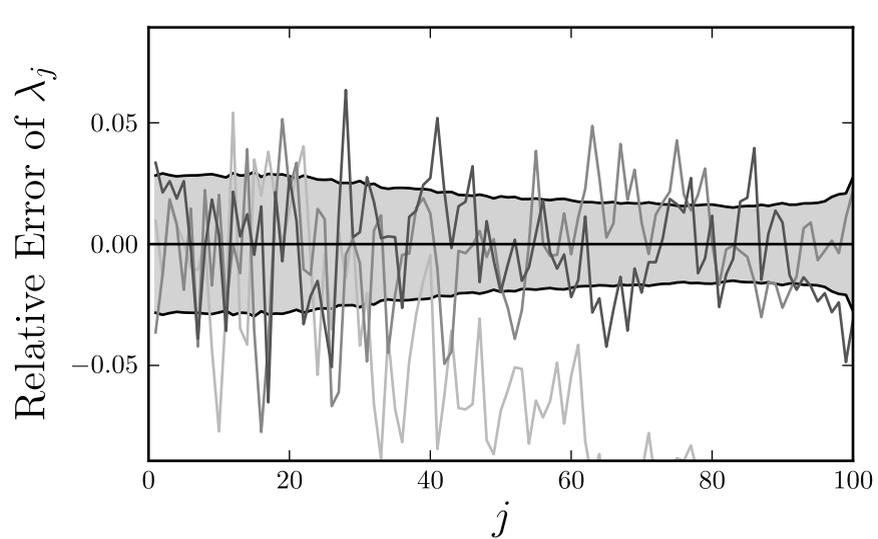
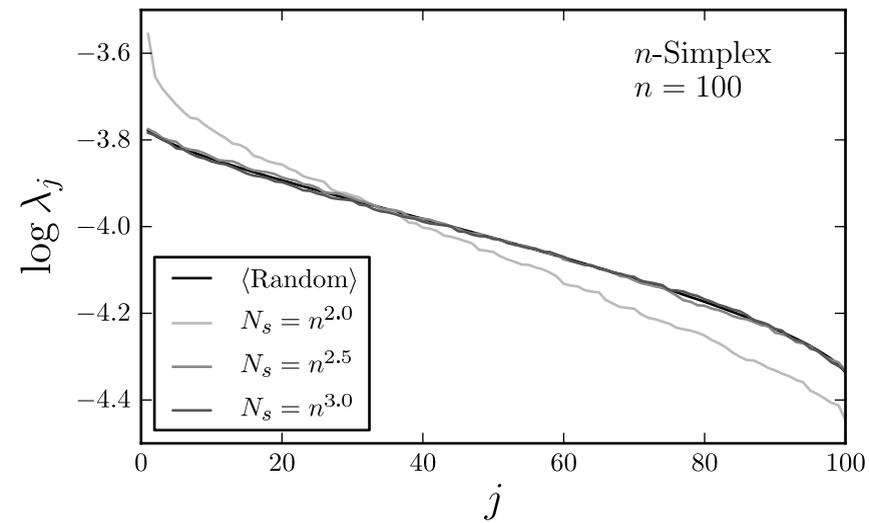
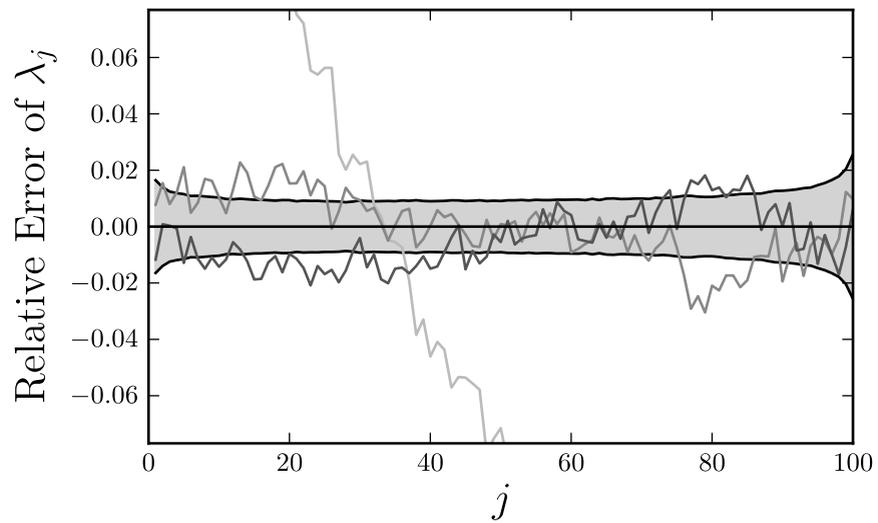
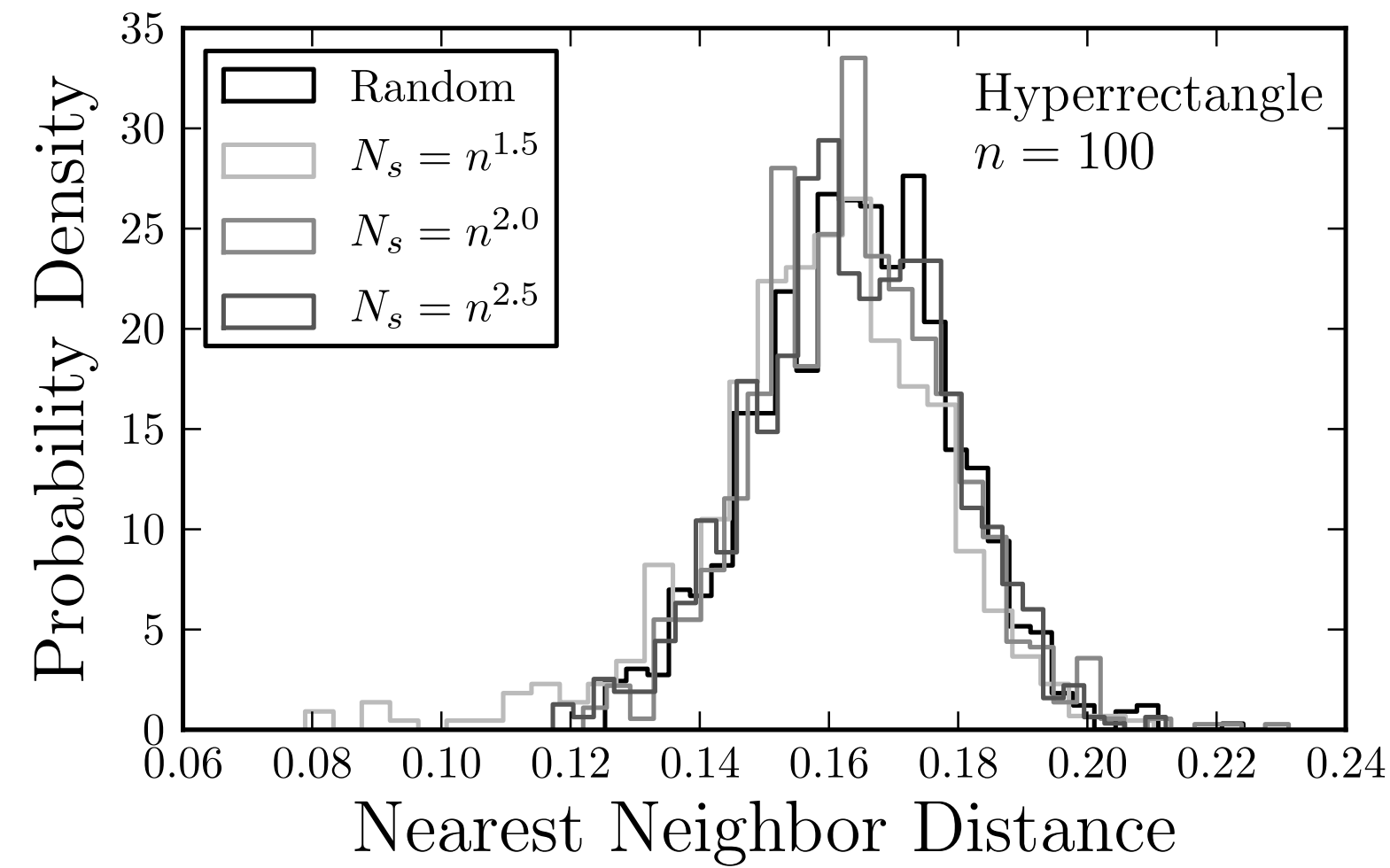
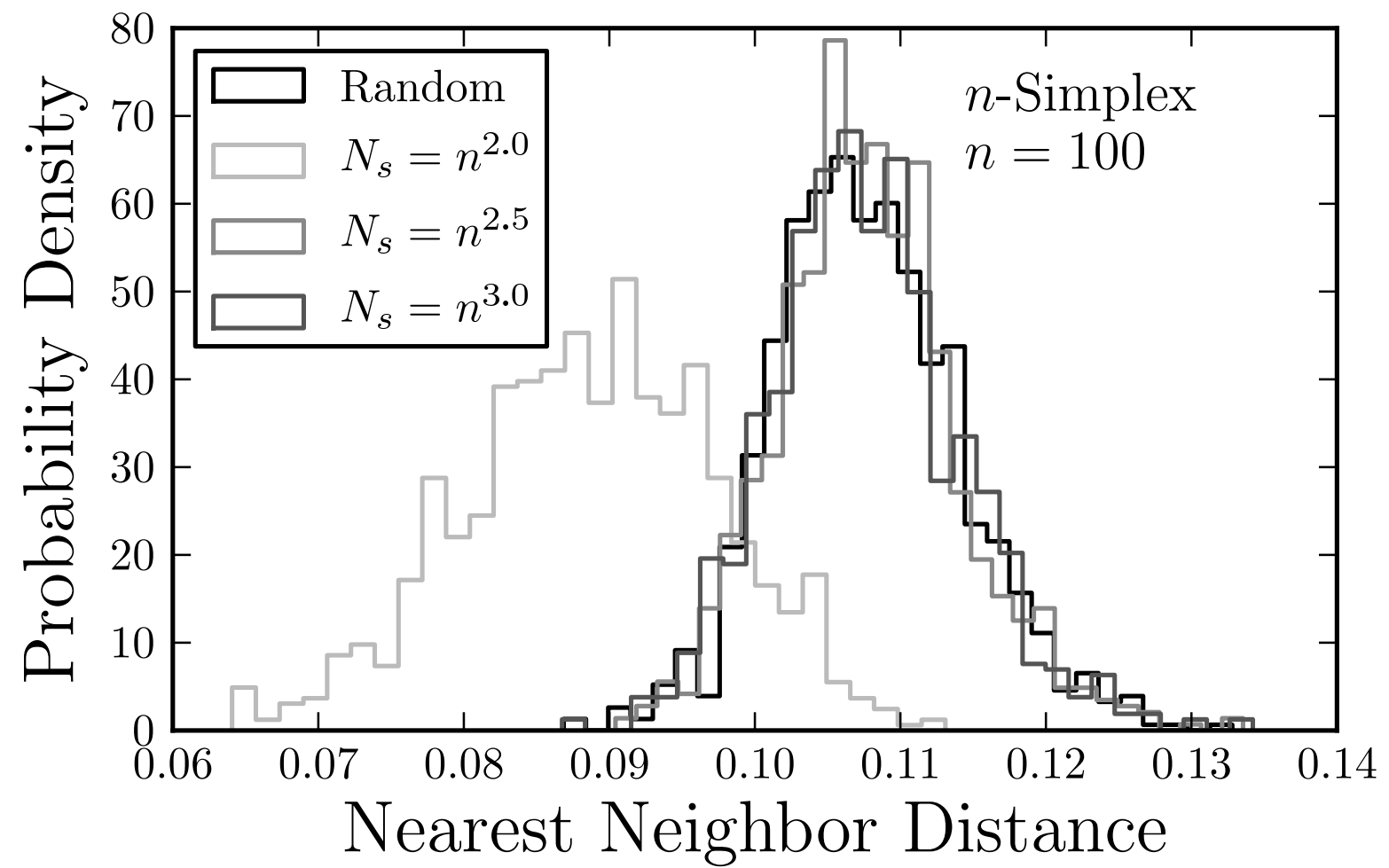
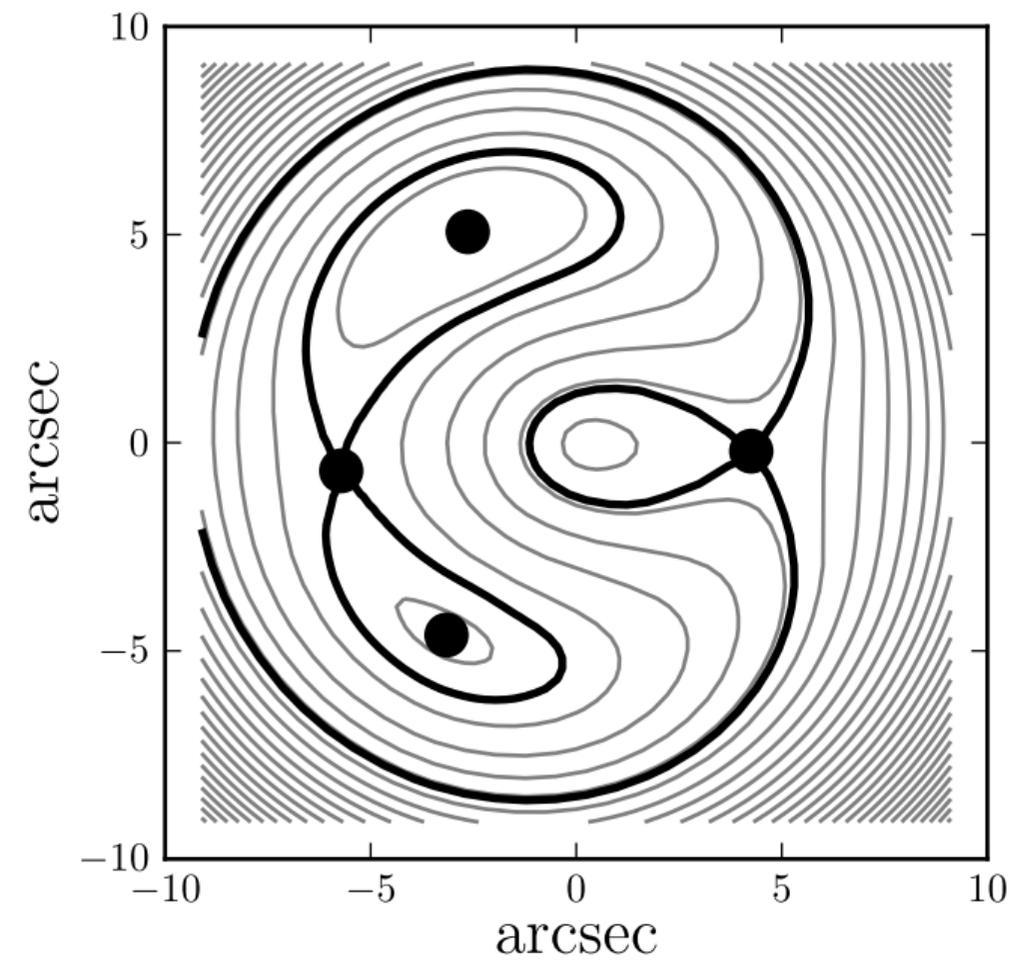
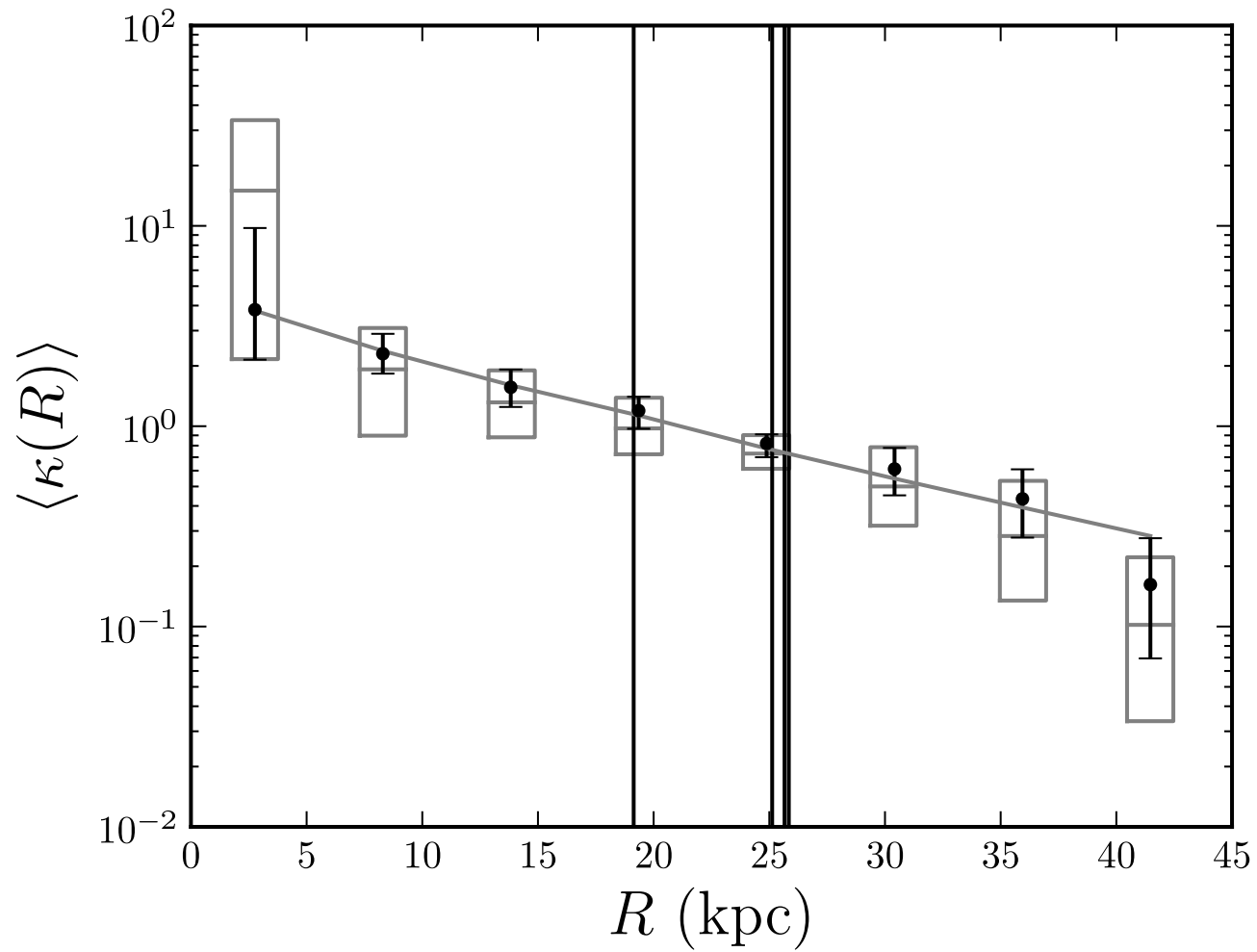
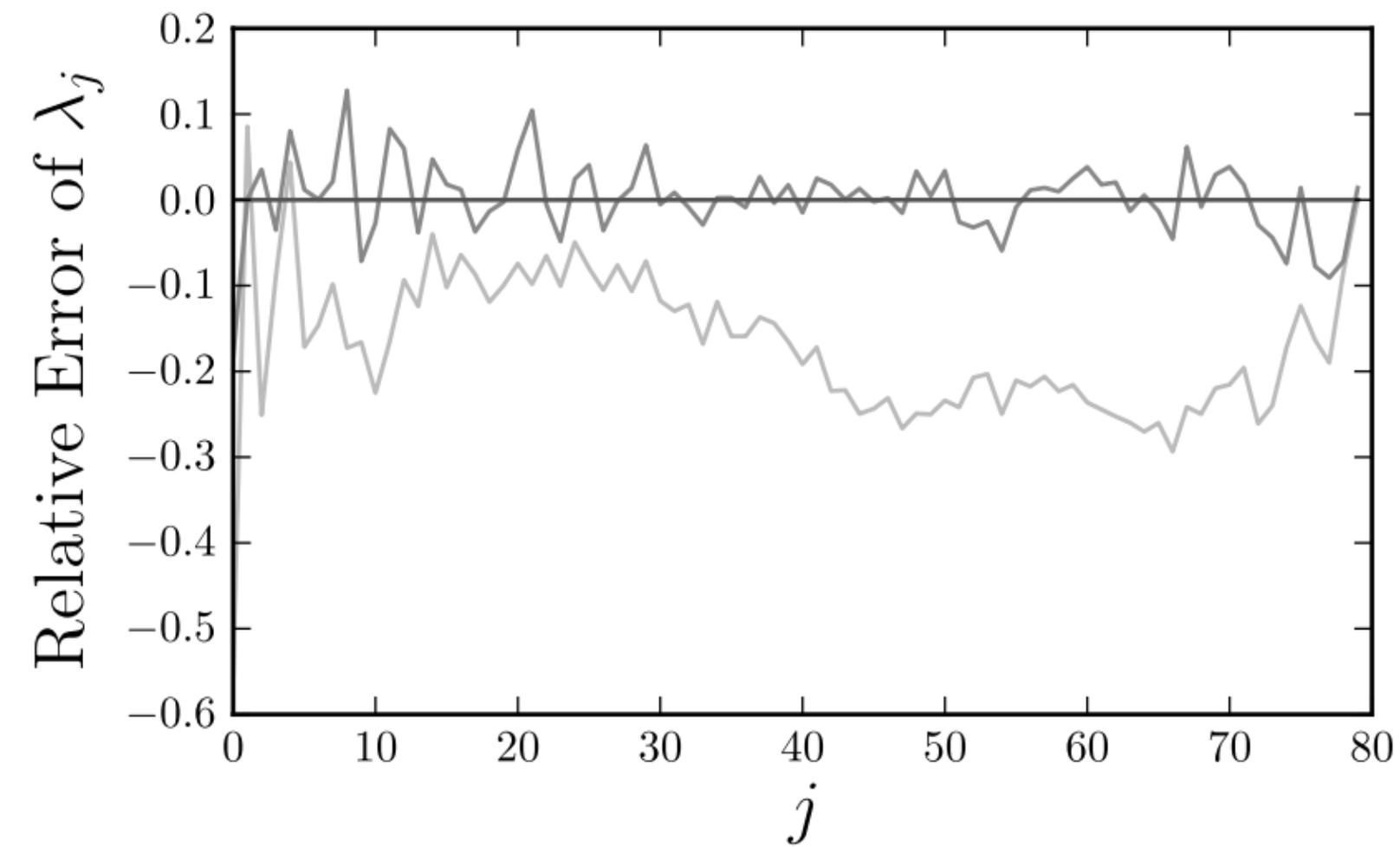
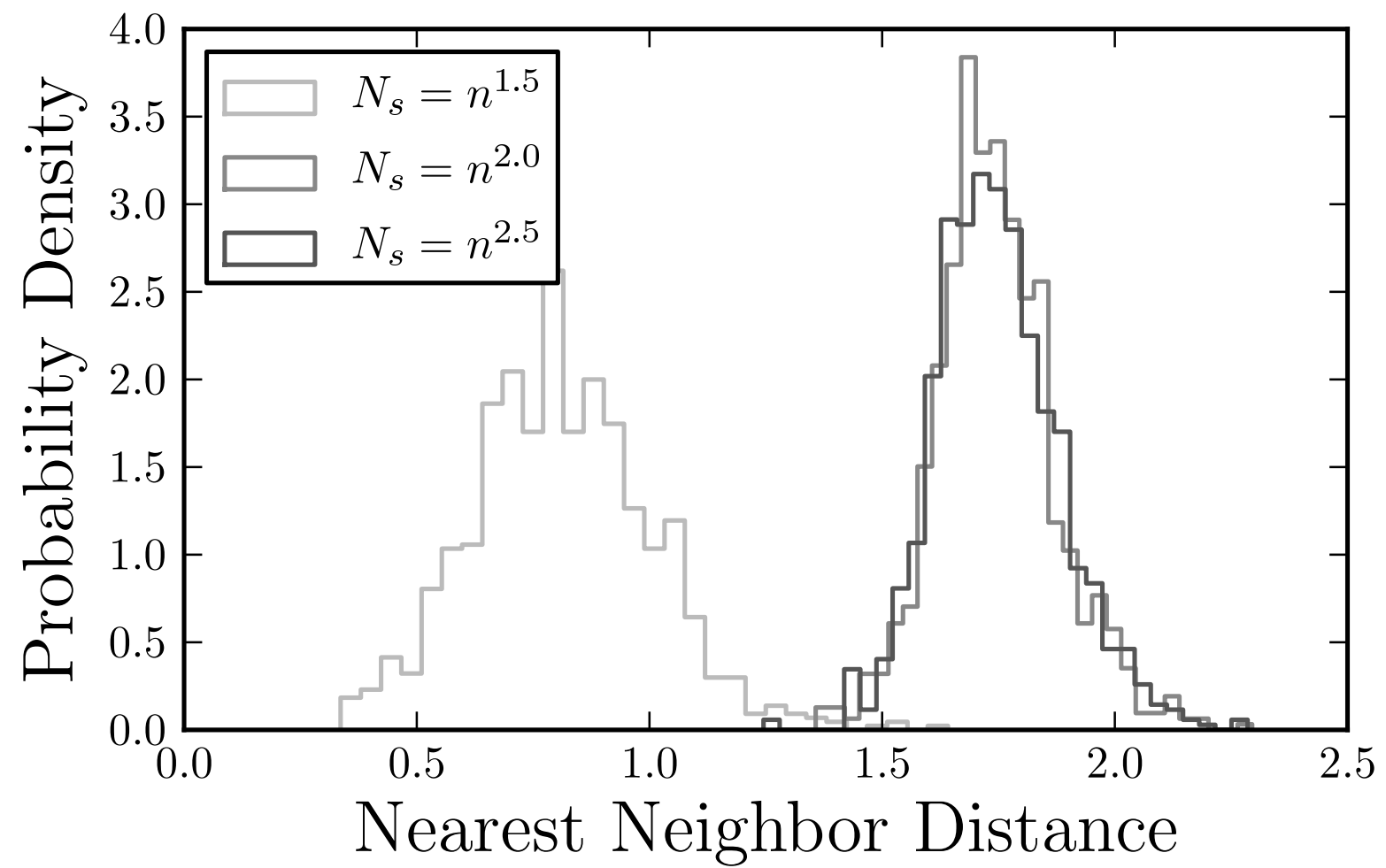
\end{document}